\documentclass[journal]{IEEEtran}
\usepackage{subfigure}
\usepackage{tikz}
\usepackage{pgfplots}
\usepackage{textcomp}
\usepackage{amsmath,amssymb,amsfonts}
\usepackage{algorithmic,tcolorbox}
\usepackage{algorithm}
\usepackage{graphicx}
\usepackage{bbding}
\usepackage{textcomp}
\usepackage{threeparttable}
\usepackage{framed}
\usepackage{verbatim}
\usepackage{bbm}
\usepackage{setspace}
\usepackage{mathrsfs}
\usepackage{bm}
\usepackage{url}
\usepackage{multirow}
\usepackage{verbatim}
\usepackage{float}
\allowdisplaybreaks[4]
\usepackage{hyperref}
\usepackage{nomencl}
\makenomenclature

\newtheorem{theorem}{Theorem}[section]
\newtheorem{lemma}[theorem]{Lemma}
\newtheorem{proposition}[theorem]{Proposition}

\newtheorem{definition}[theorem]{Definition}
\newtheorem{remark}[theorem]{Remark}

\newtheorem{assumption}[theorem]{Assumption}
\usepackage{pgfplotstable}
\pgfplotsset{
    legend image code/.code={
        \draw [#1] (0cm,-0.1cm) rectangle (0.6cm,0.1cm);
    },
}
\def\BibTeX{{\rm B\kern-.05em{\sc i\kern-.025em b}\kern-.08em
    T\kern-.1667em\lower.7ex\hbox{E}\kern-.125emX}}
\pgfplotsset{compat=1.15}

\title{Greening the Grid: Electricity Market Clearing with Consumer-Based Carbon Cost}
\author{Wenqian Jiang \emph{and} Line Roald
\thanks{W. Jiang and L. Roald are with Department of Electrical and Computer Engineering, University of Wisconsin-Madison, Madison, WI 53706 USA (e-mail: wjiang233@wisc.edu; roald@wisc.edu).}}

\begin{document}

\maketitle

\begin{abstract}
To enhance decarbonization efforts in electric power systems, we propose a novel electricity market clearing model that internalizes the allocation of emissions from generations to loads and allows for consideration of consumer-side carbon costs. Specifically, consumers can not only bid for power but also assign a cost to the carbon emissions incurred by their electricity use. These carbon costs provide consumers, ranging from carbon-agnostic to carbon-sensitive, with a tool to actively manage their roles in carbon emission mitigation. By incorporating carbon allocation and consumer-side carbon costs, the market clearing is influenced not solely by production and demand dynamics but also by the allocation of carbon emission responsibilities. To demonstrate the effect of our proposed model, we conduct a case study comparing market clearing outcomes across various percentages of carbon-sensitive consumers with differing carbon costs. Further, we conduct a comparative analysis with two strategies—carbon flow and carbon cost—to evaluate their distinct impacts on carbon emission reduction. Numerical analyses shed light on the mechanisms through which carbon costs contribute to emission reductions and inform the ongoing debate about different carbon emission reduction mechanism designs in the electricity sector.

\end{abstract}


\section{Introduction}
\label{intro}

Power generation currently stands as the leading contributor to global carbon emissions, underscoring its pivotal role in the worldwide transition towards achieving net zero emissions \cite{eia2, kayacik2024towards}. To effectively mitigate carbon emissions in power systems while simultaneously ensuring secure and affordable access to electricity for consumers, 
researchers and policy-makers have considered the introduction of
carbon taxes on electric generators \cite{green2008carbon,olsen2018optimal,algarni2020combined}. As generators roll carbon taxes into their generation costs, this strategy increases the cost of carbon-heavy generation sources and provides a competitive advantage to cleaner generators, such as solar and wind power generators. Since some generators emit more carbon than others for unit power generation, adding a carbon tax on generation might change the merit order of the generators and promote lower carbon generation dispatch solutions 
\cite{chen2023carbon, wang2020low, cheng2019low}. While this approach is relatively straightforward to implement, it has proven politically difficult to determine an adequate carbon tax \cite{metcalf2009design, harrison2010comparative}. 
Further,
it overlooks the crucial role of consumers in carbon emission reduction. However, it is the consumption of electricity that drives the need for power generation and subsequently leads to carbon emissions. Hence, it is essential to consider the impacts of consumers, especially as a growing number of them are becoming \emph{carbon-sensitive}, including data centers \cite{lindberg2021guide,Amozan, Google}, hydrogen producers \cite{zepter2023optimal, baumhof2023optimization} and individual consumers \cite{nielsen2019comparative, li2017impact}. These carbon-sensitive consumers are becoming more interested in understanding the carbon footprints of their electricity usage to optimize operations and maximize profits, highlighting the need for innovative and effective carbon emission definitions on the demand side. 
Therefore, our paper focuses on exploring consumer-centric carbon accounting and emission reduction strategies.

\begin{figure}
    \centering
    \includegraphics[width = \linewidth]{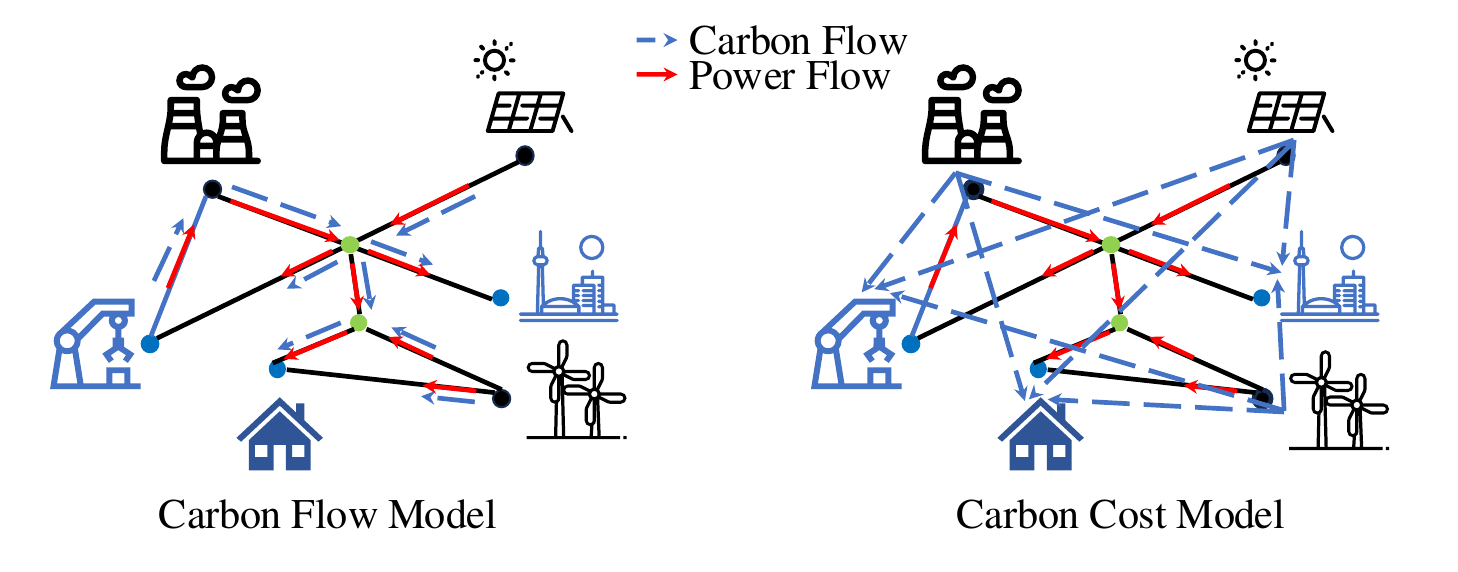}
    \caption{Comparison between carbon flow model and carbon cost model.}
    \label{fig3}
\end{figure}

There is significant ongoing debate regarding how to quantify carbon emissions for electricity consumers  \cite{miller2022hourly, hua2023demand, goetsch2022operational}. 
In practice, most companies and individuals compute their emissions based on the average rate of emissions per megawatt-hour (MWh) of electricity generated or consumed, referred to as average emissions rates \cite{kirschen1997contributions, yu2019research, li2013carbon}. The use of average emissions
complies with the Greenhouse Gas Protocol \cite{standard2014ghg}, which lists average emissions as the metric of choice for location-based accounting methods. One challenge of average carbon emissions is that this approach assumes a uniform carbon emission intensity for all consumers within a given region. 
Considering the constraints imposed by transmission line limits in real-world power systems, consumers located at various locations often exert differing, and at times conflicting, impacts on the overall carbon emission reductions \cite{ruiz2010analysis}. Consequently, average emission rates fail to create the right incentives for customers at various grid locations to contribute to emission reductions.
As a result, adapting electricity use based on average carbon emissions can in fact increase total emissions, even when it lowers the emissions assigned to any individual consumer \cite{gorka2024electricityemissions,lindberg2022using}. 
To address these drawbacks, some researchers have advocated for the use of locational marginal carbon emissions \cite{lindberg2021guide,gorka2024electricityemissions,goldsworthy2023use}, which provides information about how a change in the electricity consumption at a given node would impact total emissions. While these metrics show significant promise in effectively guiding load shifting actions by electric loads \cite{lindberg2021guide, gorka2024electricityemissions} and are currently publicly available for at least one major system operator in the United States, PJM \cite{pjm}, it also has significant drawbacks. First, the marginal emission only reflects the emission of the last generator to be dispatched, which may be higher or lower than other generators that are currently running. If we use marginal emission rates to calculate the assigned emissions for all consumers, there are typically significant differences between the total emissions from generation and the total emissions assigned to loads, meaning that marginal emissions can lead to severe over- or undercounting of emissions. Further, data from PJM also suggests that the locational marginal emissions have very high variability, several orders of magnitude more than the locational marginal prices, indicating that it might be practically challenging to use this data for load shifting. Given the drawbacks of these existing carbon emission metrics, recent research has explored new metrics, such as adjusted locational marginal emissions \cite{gorka2024electricityemissions} or locational average emissions based on carbon flow \cite{chen2023carbon}.

The Greenhouse Gas Protocol outlines a second option for carbon accounting of electricity, referred to as market-based accounting. In this method, consumers can purchase renewable energy certificates (RECs) \cite{USEP,EECS, feng2024wind, hulshof2019performance}, often coupled with power purchase agreements (PPAs) \cite{bruck2018levelized,brunnberg2019power}, to claim that they are carbon-free. This method decouples physical electricity consumption from carbon emissions accounting, as consumers may utilize carbon-intensive electricity without punishment as long as they procure a comparable amount of RECs. Clearly, this approach, which currently is based on yearly consumption, cannot provide real-time signals for consumers to optimize their electricity consumption patterns and reduce carbon emissions. Developing regulations in the United States \cite{USEP} and Europe \cite{EUEP} seek to establish new systems where RECs or comparable certificates are issued on an hourly basis. To claim that their electricity is green, consumers will have to match their hourly consumption with concurrently generated renewable energy in the same region where they are located. This promises to tie the carbon accounting more closely to physical electricity use, though some challenges will persist.
In particular, in both current location-based and market-based carbon accounting, the accounting step is separate from the market clearing. As a result, consumers only get to know the values of the carbon metrics, or amount of available RECs, \emph{after} they have consumed their energy. This requires consumers to accurately predict carbon emissions and RECs availability ahead of time, which can be a challenging task. A more appealing and efficient option is to internalize carbon emissions in the market clearing. 
In their recent work, \cite{chen2023carbon, kang2015carbon} leveraged the carbon flow method to assign carbon emissions to consumers and defined a market clearing mechanism where consumers can place explicit limits on the carbon emissions they are willing to incur.
The core concept of carbon flow is that the carbon emissions are virtually attached to the power flowing from generators to loads. To enable the tracing of the power flow from source to sink, the carbon flow method makes the assumption that power flowing into a node is split between the outgoing power flows according to the proportional sharing principle \cite{kirschen1997contributions, bialek1996tracing}, as depicted in Fig. \ref{fig3}. Based on this assumption about power tracing, the carbon intensity for each node where consumers are located can be calculated and, as is done in \cite{chen2023carbon}, explicitly limited. One challenge of the carbon flow method is that it is not clear that the proportional sharing principle is the “right" definition of the carbon flow tracing. 
Furthermore, it can be very challenging for consumers or the independent system operator (ISO) to define what the values of the carbon emission cap should be.  
Despite the latter issue of defining carbon limits, other researchers have also explored this direction \cite{wang2021optimal, pietzcker2021tightening}. Imposing carbon limits on the load requires that consumers curtail their electricity consumption once predefined emission levels are exceeded.  
Even if consumers are good at setting the carbon emission cap, they still lack strong incentives to adhere to it for emission reduction purposes. 

In this paper, we propose a new market clearing method that allocates carbon emissions from generation to consumers by optimally allocating specific amounts of power from each generator to each load. This allows us to “trace" power in an optimal way, without making restrictive assumptions that are grounded in physics. Further, we propose that it is more natural for consumers to assign a cost to carbon emissions, rather than defining carbon emission limits. Our method therefore assumes that consumers have the opportunity to submit non-negative carbon cost information along with their bids to consume electricity. 
Our market clearing formulation draws inspiration from practices related to renewable energy certificates and the idea behind multi-commodity network flow problems, 
as illustrated in the carbon cost model in Fig. \ref{fig3}.

The major contributions of this paper are twofold. First, we propose a novel electricity market clearing model with consumer-based carbon cost, which optimally allocates generation from different generators to different consumers. This novel formulation provides a new perspective to both account for carbon emissions of end-users and reduce carbon emissions of the power system. To the best of our knowledge, this is the first paper that investigates carbon footprints and carbon reductions in power systems by involving carbon costs submitted by consumers. 
Second, we demonstrate how the proposed model impacts market clearing outcomes across a range of different scenarios and against a variety of different benchmarks.
Our extensive numerical study includes results both for the simplified three-bus system and the real-world IEEE RTS-GMLC system. We first compare our proposed model with benchmark models with or without considering the impact of carbon emissions. 
We then investigate how different percentages of carbon-sensitive consumers and varying carbon costs impact the generation dispatch, consumption level, and emissions. 

The remainder of the paper is organized as follows. Section \ref{sec2} describes our proposed market clearing model, while Section \ref{sec3} provides details of our benchmark formulations. In Section \ref{ns}, we show results and conclusions from our numerical study, while Section \ref{conclusion} summarizes and concludes.

\section{Market Clearing with Price and Carbon Information}
\label{sec2}
\subsection{Motivation}
To enhance carbon emission reduction efforts in power systems, 
we propose a novel electricity market clearing model that (i) incorporates the allocation of carbon emissions from electricity generation to loads into the market clearing itself and (ii) allows for consideration of consumer-side carbon costs. There are two main reasons that motivated us to design this new model. First, current location-based and market-based demand-side carbon accounting separate carbon emission allocation from the electricity market clearing (ususlly after generation dispatch decisions are being made), which brings challenges for both the ISO and consumers as discussed above. 
In contrast, we propose that the carbon emission responsibilities for each consumer can be allocated internally in the electricity market. This provides a new perspective on utilizing market clearing to achieve fair and equitable carbon allocation.   
Second, it is clear that not all consumers are inclined to shoulder the responsibility for reducing carbon emissions. On the one hand, carbon-agnostic consumers lack the incentive to participate in emission reduction efforts. On the other hand, an increasing number of carbon-sensitive entities, such as data centers and hydrogen generation facilities, have strong incentives to ensure that their electricity consumption is as environmentally friendly as possible. Considering the diverse attitudes of consumers towards carbon reduction, we introduce a novel mechanism by which consumers can submit information about their cost of carbon emissions from electricity use. 
This mechanism is compatible with the established bidding mechanisms for power in traditional electricity markets, as consumers are not required to submit carbon costs, but can choose to do so. 


\subsection{Problem Formulation}
We consider an electric power network with the set of nodes, consumers, transmission lines, and generators denoted by $\mathcal{N}$, $\mathcal{D}$, $\mathcal{L}$ and $\mathcal{G}$, respectively. Let $\mathcal{G}_i\subset \mathcal{G}$ and $\mathcal{D}_i\subset \mathcal{D}$ be the subset of generators and loads connected to node $i$, and $(i,j)\in\mathcal{L}$ denote the transmission line from node $i$ to node $j$. 
Prior to the market clearing, each generator submits their generation costs $c_{g,m}$, and maximum and minimum generation capacities $P_{g,m}^{\max}$ and $P_{g,m}^{\min}$. Each consumer submits their bids for consumption $u_{d,n}$, reflecting the utility (or revenue) they derive from consuming electricity, and information about their maximum and minimum demand $P_{d,n}^{\max}$ and $P_{d,n}^{\min}$. Moreover, each consumer also submits information about their carbon cost $c_{co_2,n}$, given in units of [\$/tons $CO_2$]. This cost may be directly tied to concrete costs such as carbon emission penalties from carbon taxes or cap-and-trade schemes, or could be an internally set “carbon cost", reflecting how much revenue the consumer is willing to forgo to avoid carbon emissions. An interesting attribute of the proposed scheme is that the carbon costs are determined directly by the consumers, and would be likely to vary between different groups of customers. Importantly, carbon-agonostic consumers can choose to set their carbon cost to zero, i.e. $c_{co_2,n}=0$, and would not have to provide any additional information beyond what is submitted to the current market clearing. 

Once bids for generation, consumption, and carbon are known, the ISO solves the market clearing problem. This problem is a modified version of the optimal power flow (OPF) problem, where we have adapted the objective function to consider consumers' carbon cost and added constraints to assign generated power (and associated emissions) from each generator to each load. The problem can be formulated as the following optimization problem: 
\begin{align}
    \max_{P_g, P_d, \theta, \pi, E_d}&\ u_d^\intercal P_d-c_{co_2}^\intercal E_d -c_g^\intercal P_g\label{eq1coo}\\
    s.t.
    & \sum_{n\in \mathcal{D}_i}P_{d,n}+\sum_{j:(i,j)\in \mathcal{L}}\beta_{ij}(\theta_i-\theta_j)=\sum_{m\in \mathcal{G}_i}P_{g,m},\notag \\
    &\forall i \in \mathcal{N},\tag{\ref{eq1coo}{a}} \label{eq1coa}\\
    &-F_{ij}^{\rm{lim}}\leq\beta_{ij}(\theta_i-\theta_j)\leq F_{ij}^{\rm{lim}}, \quad\forall (i,j)\in \mathcal{L},\tag{\ref{eq1coo}{b}} \label{eq1cob}\\
    &P_{g,m}^{\min}\leq P_{g,m}\leq P_{g,m}^{\max}, \quad\forall m \in \mathcal{G} \tag{\ref{eq1coo}{c}} \label{eq1oc},\\
    &P_{d,n}^{\min}\leq P_{d,n}\leq P_{d,n}^{\max},\quad\forall n \in \mathcal{D}, \tag{\ref{eq1coo}{d}}\label{eq1od}\\
    &\theta_{\rm{ref}} = 0\tag{\ref{eq1coo}{e}} \label{eq1coe},\\
    & \sum_{n\in\mathcal{D}}\pi_{m,n} = P_{g,m},\ \forall m \in \mathcal{G}, \tag{\ref{eq1coo}{f}} \label{eq1cof}\\
    & \sum_{m\in\mathcal{G}}\pi_{m,n} = P_{d,n},\ \forall n \in \mathcal{D},\tag{\ref{eq1coo}{g}} \label{eq1cog}\\
    & \sum_{m\in\mathcal{G}}e_{g,m}\pi_{m,n} = E_{d,n}, \ \forall n \in \mathcal{D} \tag{\ref{eq1coo}{h}} \label{eq1cho},\\
    & \pi_{m,n}\geq 0,\ \forall m \in \mathcal{G}, \ \forall n \in \mathcal{D} \tag{\ref{eq1coo}{i}} \label{eq1cooi}.
\end{align}
Here, the optimization variables are the generation dispatch $P_g = \{P_{g,m}: m \in \mathcal{G}\}$, the voltage angle $\theta = \{\theta_i: i\in \mathcal{N}\}$, the flexible load $P_d = \{P_{d,n}:n \in \mathcal{D}\}$, the generation-load allocation matrix $\pi=\{\pi_{m,n}:m \in \mathcal{G}, n \in \mathcal{D}\}$ reflecting the amount of power assigned from each generator to each load, and the total carbon emission for each consumer $E_d = \{E_{d,n}:n\in \mathcal{D}\}$. The objective function (\ref{eq1coo}) maximizes social welfare considering cost of generation $c_g$, consumer utility derived from electricity consumption $u_d$,  and consumer carbon costs $c_{co_2}$. 

The constraints (\ref{eq1coa})-(\ref{eq1coe}) are similar to those of a standard DC OPF.
Constraint (\ref{eq1coa}) ensures that nodal power balance constraints are met, with $\beta_{ij}\in \mathbb{R}$ denoting the susceptance value of the transmission line $(i,j)$ from node $i$ to node $j$. Constraints (\ref{eq1cob}) are the transmission line limits, where $F_{ij}^{\rm{lim}}$ represents the transmission capacity (which we assume is the same in both directions). Constraints (\ref{eq1oc}) and (\ref{eq1od}) ensure that limits on generation capacity and demand flexibility are enforced, while constraint (\ref{eq1coe}) sets the voltage angle at the reference node to zero.  

The remaining constraints (\ref{eq1cof})-(\ref{eq1cooi}) represent the mechanism for allocating carbon emissions from generation to loads. This mechanism assigns a fraction of the power from each generator to each load, and then computes the total load emissions $E_d$ based on emission intensity and the amount of power obtained from each generator.
To facilitate the allocation of electricity from each generator to each consumer,  
we introduce the generation-load allocation pair matrix, denoted by $\pi$. 
Constraint (\ref{eq1cof}) ensures that the total amount of power allocated from the generator $m$ to all the loads $n\in\mathcal{D}$ equals the actual power dispatched from this generator. Constraint (\ref{eq1cog}) similarly enforces that the sum of all allocations of power to a given load is equal to the total load power consumption. 
Constraint (\ref{eq1cho})
calculates the carbon emissions $E_{d,n}$ for each consumer, where $e_{g,m}$ denotes the carbon emission intensity of generator $m$. Constraint (\ref{eq1cooi}) ensures that all load allocations are non-negative, which ensures that all loads will have non-negative emissions, assuming non-negative generator emission intensities.

\textit{Remark 1}: The consumer-based carbon cost in Problem (\ref{eq1coo}) provides freedom for consumers to choose if they are willing to pay for carbon emissions or not. That is, if a consumer $n$ is unconcerned about carbon emissions, they may submit a carbon bid $c_{co_2}=0$, resulting in a higher carbon emission allocation for that consumer. Otherwise, if the consumer is carbon-sensitive, submitting higher carbon costs will lead to lower assigned carbon emissions. 

Note that we do not necessarily envision the term $c_{co_2}^\intercal E_d$ in the objective function to be associated with any payment from the ISO to the load. The carbon cost $c_{co_2}^\intercal E_d$ is intended to reflect how a consumer's utility of consuming electricity is reduced when the consumed electricity comes from carbon-emitting sources. However, consumers are expected to pay for electricity according to their provided consumption bids $u_d$, even if a non-zero carbon cost is incurred. The benefit of submitting a non-zero carbon cost to the electricity market is that the market clearing with constraints (\ref{eq1coo}{f})-(\ref{eq1coo}{i}) will allocate more low-carbon power to the consumers with the highest carbon costs. However, if a limited amount of low-carbon power is available, the consumers with the high carbon cost may also be the first to be dispatched at reduced consumption, as their carbon costs will start competing with the utility of consuming electricity.

\section{Benchmark Formulations}
\label{sec3}
In this section, we present three benchmark models for comparative analysis. These models comprise two carbon-agnostic frameworks, with and without consideration of demand flexibility, alongside one carbon-aware formulation, the carbon flow model.
\subsection{Standard (Carbon-Agnostic) Formulations}
 \subsubsection{Market Clearing with Fixed Demands}
 The traditional electricity market solves the market clearing problem given a fixed demand (e.g., day-ahead market). For simplicity, we assume that the demand of each consumer is fixed at their maximum values. Mathematically, the problem is defined as
\begin{align}
    \min_{P_g, \theta}&\ c_g^\intercal P_g\label{eq123}\\
    \nonumber
    s.t.\ 
    & \sum_{n\in \mathcal{D}_i}P_{d,n}^{\max}+\sum_{j:(i,j)\in \mathcal{L}}\beta_{ij}(\theta_i-\theta_j)=\sum_{m\in \mathcal{G}_i}P_{g,m},\notag \\
    &\forall i \in \mathcal{N},\tag{\ref{eq123}{a}} \label{eq123a}\\
    & \rm{Constraints} \ (\ref{eq1cob}), \ (\ref{eq1oc}), \ (\ref{eq1coe})\notag.
\end{align}

 \subsubsection{Market Clearing with Demand Flexibility}
 We next consider the case in which consumers have flexibility in their electricity consumption, similar to what we assumed in our proposed model, but do not have the ability to submit information about their carbon costs. 
In this case, the optimization problem can be defined as follows:
 \label{app4}
 \begin{align}
    \max_{P_g, P_d, \theta}&\ u_d^\intercal P_d - c_g^\intercal P_g\label{eq1234}\\
    \nonumber
    s.t.\ 
    & \rm{Constraints} \ (\ref{eq1coa})-(\ref{eq1coe})\notag.
\end{align}
We note that this model is equivalent to a version of the proposed market clearing model \eqref{eq1coo} where all loads submit carbon costs $c_{co_2}=0$. This is because there will always exist a feasible allocation of generation to loads, and if all loads have zero carbon costs, the emission allocation would have no impact on the market clearing outcome. 

\subsection{Carbon Flow Model}
\label{app1}
This model formulation is based on the concept of carbon emission flow proposed in \cite{kang2015carbon, chen2023carbon}. There are two critical fundamental principles of this model: 1) Similar to the nodal power balance constraint, the total carbon inflows equal the total carbon outflows at each node, i.e., the nodal carbon mass is preserved; 2) The allocation of total carbon inflows to the different carbon outflows is proportional to their power flow values at each node. This principle of sharing power (and associated emissions) is referred to as the proportional sharing principle \cite{kirschen1997contributions}. The original carbon flow model adopts carbon emission limit settings for emission reductions. To make it more comparable to our proposed model, we use a model that incorporates carbon costs and the carbon flow constraints, giving rise to the following optimization problem:
\begin{align}
    \max_{P_g, P_d, \theta, \lambda_e}&\ \ u_d^\intercal P_d-c_{co_2}^\intercal E_d -c_g^\intercal P_g\label{eq1}\\
    \nonumber
    s.t.\ 
    & \lambda_{e,i}\cdot \left(\sum_{n\in \mathcal{D}_i}P_{d,n}+\sum_{j,(i,j)\in \mathcal{L}}\beta_{i,j}(\theta_i-\theta_j)\right)\\
   &=\sum_{m\in \mathcal{G}_i}e_{g,m}P_{g,m}+\sum_{j,(i,j)\in \mathcal{L}}\lambda_{e,j}\cdot \beta_{i,j}(\theta_j-\theta_i), \notag\\
   & \forall i \in \mathcal{N}, \tag{\ref{eq1}{a}} \label{eq1a}\\
    & \lambda_{e,i}\cdot P_{d,n} = E_{d,n}, \ \forall i \in \mathcal{N}, \ \forall n \in \mathcal{D}_i \tag{\ref{eq1}{b}} \label{eq1b},\\
    & \rm{Constraints} \ (\ref{eq1coa})-(\ref{eq1coe})\notag.
\end{align}
Here, $\lambda_{e,i}$ is the carbon intensity for the node $i$ and the constraint (\ref{eq1b}) calculates carbon emissions for each consumer. We note that this problem has bilinear terms in the constraint (\ref{eq1a}), where both $\lambda_e$ and $\theta$ or $P_d$ are variables. Thus Problem (\ref{eq1}) is a non-convex problem, and there is a chance that any optimal solution is only locally optimal rather than global optimal. 

A comparison with our proposed model is appropriate. First, the carbon flow-based formulation (\ref{eq1}) is a non-convex problem, whereas the proposed model \eqref{eq1coo} is linear and convex. Second, problem (\ref{eq1})
imposes specific constraints on how emissions from generators are allocated to loads. 
Our proposed model has more flexibility in how generation and carbon emissions are allocated from generators to loads, and only enforces power balance for the individual generators and loads. As a result, any carbon emission allocation that the carbon flow model proposes would also be a feasible allocation for our model. However, our model has the ability to choose a more ``optimal'' allocation of the carbon emissions, such that more low-carbon generation is allocated to the loads with the higher carbon costs.




\section{Numerical Studies}
\label{ns}
 This section conducts extensive numerical studies to demonstrate how our proposed carbon cost model could impact electricity market clearing. All the tests are performed on a laptop computer with an Apple M3 Pro CPU and 36GB RAM. The optimization problem is solved using both GAMs \cite{GAMs} and Julia \cite{bezanson2017julia}.

 
\subsection{Evaluation Metrics}
In our case study, we analyze the optimal solutions of different instances and problem formulations to understand how different aspects of the market outcome change. Specifically, we consider the following quantities when evaluating our solutions: 
\begin{itemize}
    \item Generation dispatch ($P_g$) and demand allocation ($P_d$);
    \item Carbon emission responsibility allocation for each consumer (defined as $E_d$, $\sum_{m}e_{g,m}\pi_{m,n}$, or $\lambda_{e,i}\cdot P_{d,n}$, depending on the formulation); 
    \item Total generation: $\sum_{m\in \mathcal{G}}P_{g,m}$;
    \item Total generation cost: $\sum_{m\in \mathcal{G}}c_{g,m}P_{g,m}$;
    \item Total carbon: $\sum_{m\in \mathcal{G}}e_{g,m}P_{g,m}$;
    \item Average carbon: Total carbon / Total generation. 
\end{itemize}
\subsection{Simplified Three-bus System}
\begin{figure}
    \centering
    \includegraphics[width = 0.68\linewidth]{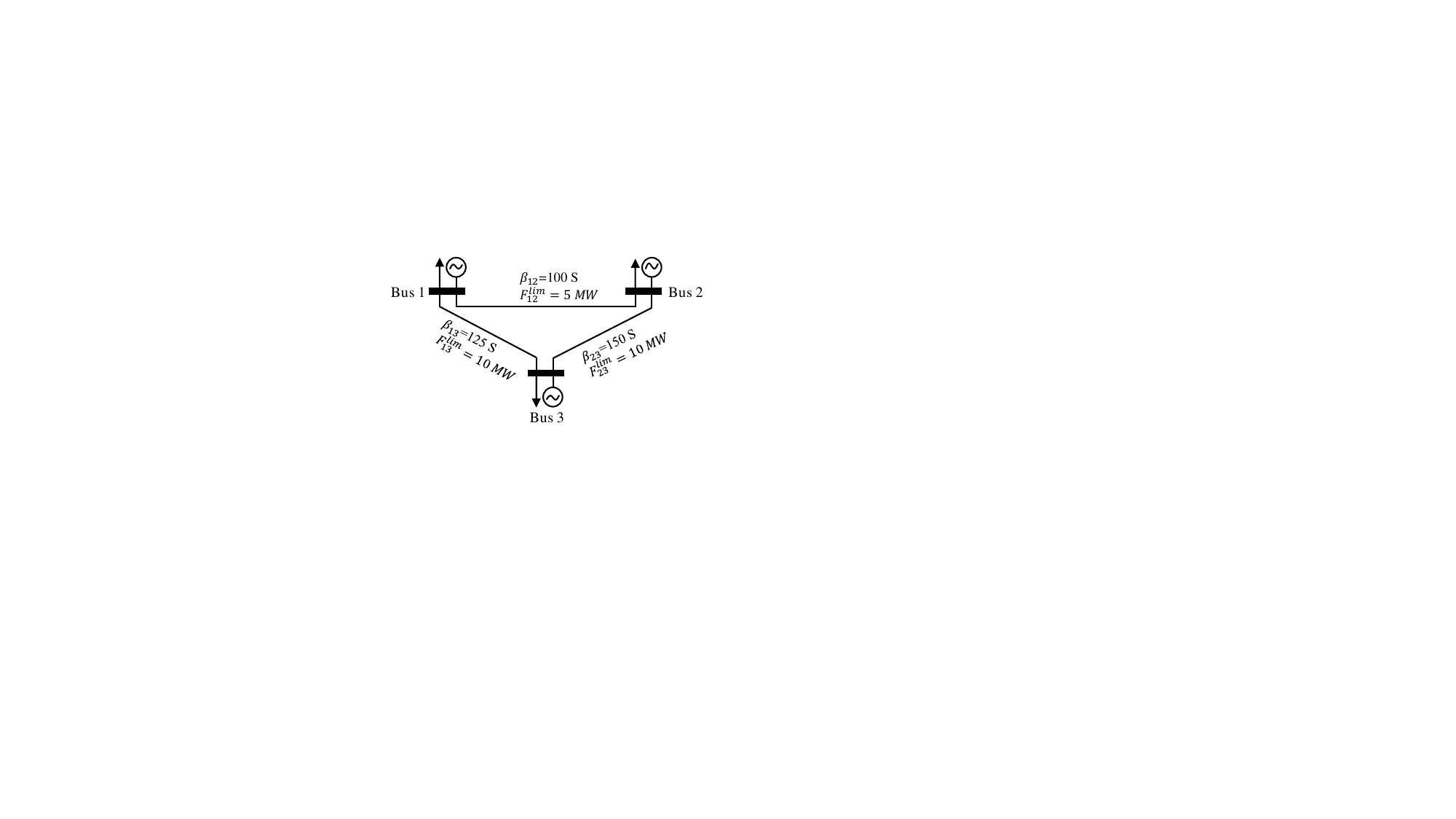}
    \caption{Illustration of the three-bus system.}
    \label{fig1}
\end{figure}

To be able to describe our results in detail, we first consider a simple three-bus system adapted from Example 6.2.2 in \cite{gabriel2012complementarity}. The network is shown in Fig. \ref{fig1} which includes three generators and three consumers, with one generator and consumer located at each bus. Since the carbon intensity of the generators is not included in the original three-bus system data, we define different carbon intensities for all generators. We assume that the most expensive generator, located at Bus 2, has the highest carbon intensity, while the cheapest generator, located at Bus 3, has the lowest carbon intensity. The medium cost generator at Bus 1 has an intermediate carbon intensity. 
All system parameters are provided in Table \ref{tab2} and Fig. \ref{fig1}. 

\begin{table}
\footnotesize
\centering
\caption{Parameters of Generators and Consumers.}
\label{tab2}
\begin{tabular}{ccccc}
\hline
&Bus (\#)& 1&2 &3\\
\hline
\multirow{3}{*}{Consumers}&$P_{d}^{\rm{min}}$(MW)&4&16&12\\
&$P_{d}^{\rm{max}}$(MW)&6&24&18\\
&$u_d$(\$/MWh)&18&20&21\\
\hline
\multirow{3}{*}{Generators}&$P_{g}^{\rm{min}}$(MW)&0&0&0\\
&$P_{g}^{\rm{max}}$(MW)&20&10&25\\
&$c_g$(\$/MWh)&8&10&6\\
&$e_g$(tons/MWh)&0.6&1&0.2\\
\hline
\end{tabular}
\end{table}

\begin{figure}[!htpb]
\centering
\subfigure[Individual level comparison.]{
\includegraphics[width=0.95\linewidth]{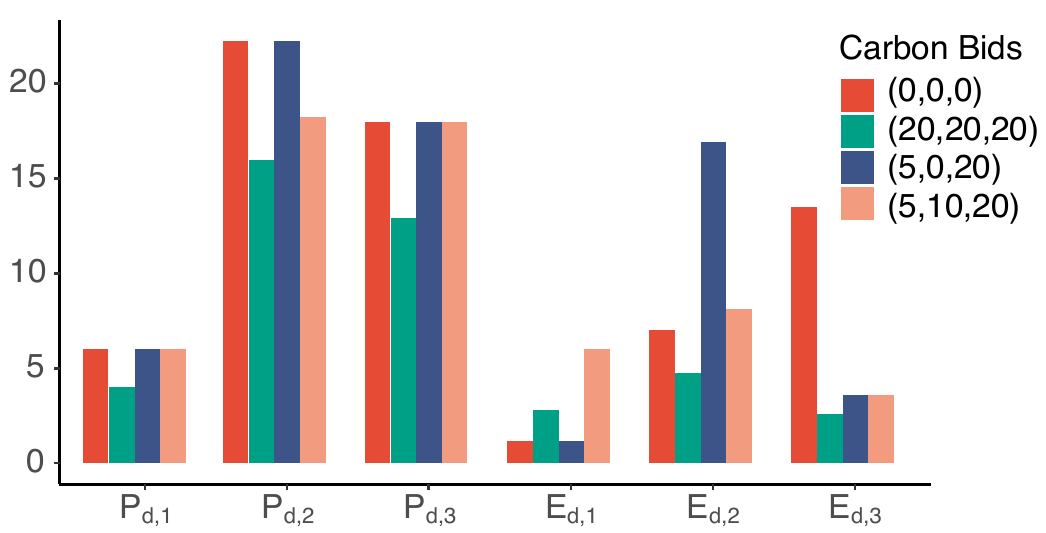}
}

\subfigure[System level comparison.]{
\includegraphics[width=0.92\linewidth]{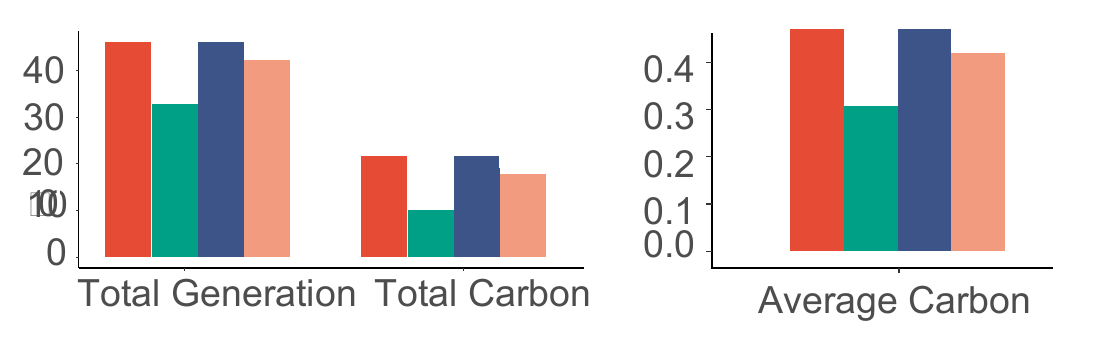}
}
\caption{Results for different carbon costs in the three-bus system. }
\label{fig2}
\end{figure}

Based on the above system parameters, we analyze how different carbon costs $c_{co_2}$ impact optimal solutions of Problem (\ref{eq1coo}). We consider four different bids, listed in the legend of Fig. \ref{fig2}. Fig. \ref{fig2} also shows the results for total generation, total carbon emissions, electricity consumed by each consumer, and total emissions allocated to each consumer. 

We first compare the case without carbon costs and the case with carbon costs (red bars and green bars) in Fig. \ref{fig2}. After involving carbon costs, the total generation reduces 28.8\% while the total carbon emission reduces more than half (53.3\%), highlighting significant carbon mitigation performances. An interesting observation is that although carbon costs drive all consumers to reduce their demands (all are equal to or close to minimum load values), the carbon emission allocated to the consumer $1$ actually increases. This unexpected result is explained by the fact that the carbon emission allocation for the case with zero bids (red bars) is arbitrarily defined by the optimization problem, since none of the consumers assign any cost to the emissions.  

We next compare the case with carbon costs $(5,0,20)$ and the case with carbon bids $(5,10,20)$ (blue bars and pink bars). In this case, only the consumer $2$ changes the carbon bid from $0$ to $10$. Although amounts of power consumed at bus $1$ and $3$ do not change, the carbon emission allocation to those loads changes a lot, as emissions are shifted from the load at bus 2 (previously the load with the lowest carbon cost) to the load at bus 1 (now the load with the lowest carbon cost). 

\subsection{IEEE RTS-GMLC System}

\begin{figure}[!htpb]
    \centering
    \includegraphics[width = \linewidth]{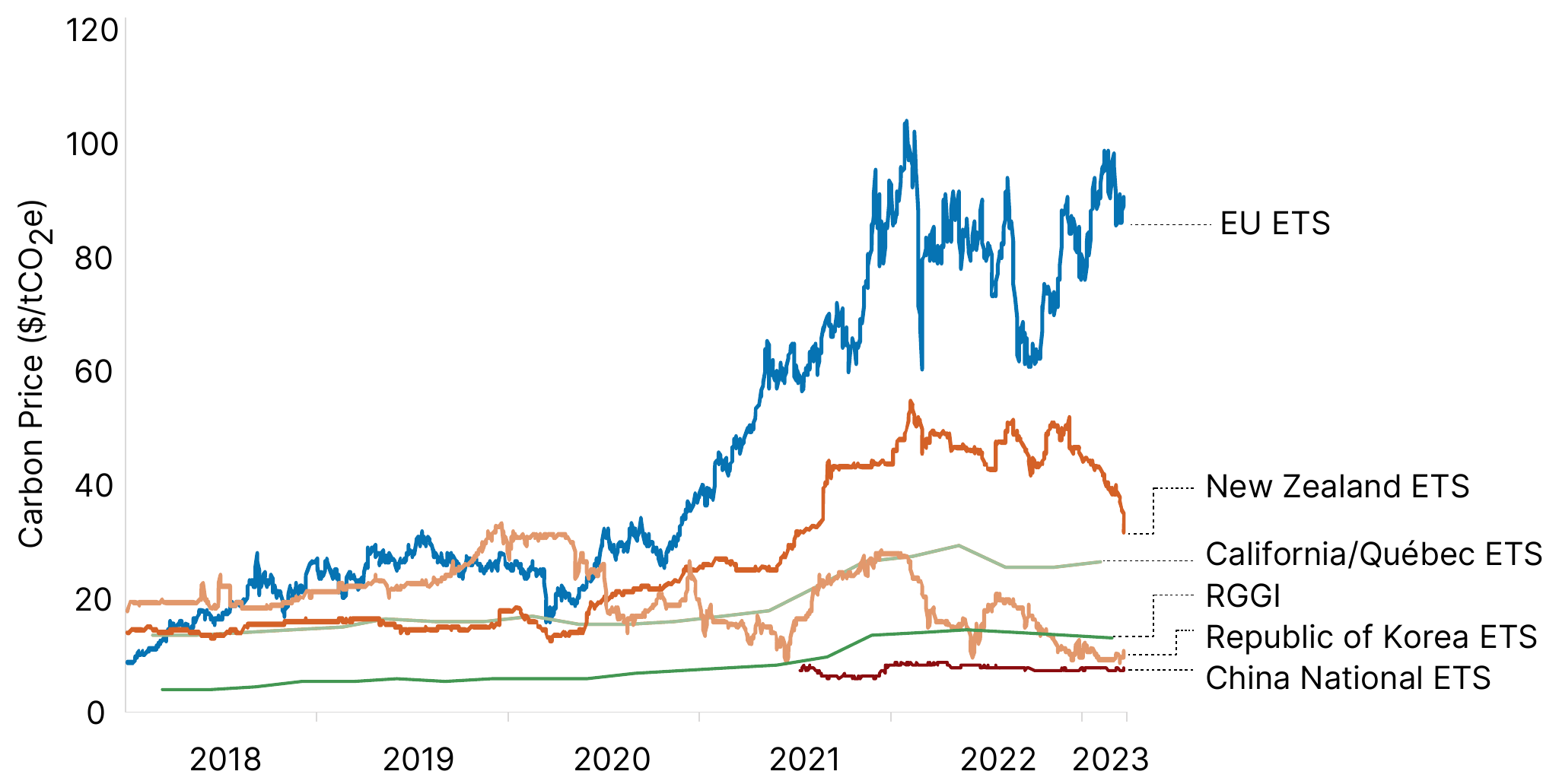}
    \caption{Carbon prices in several emission trading markets from 2018 to 2023 \cite{carbonp}.}
    \label{figcp}
\end{figure}

\begin{figure}[!htpb]
    \centering
    \includegraphics[width = 0.95\linewidth]{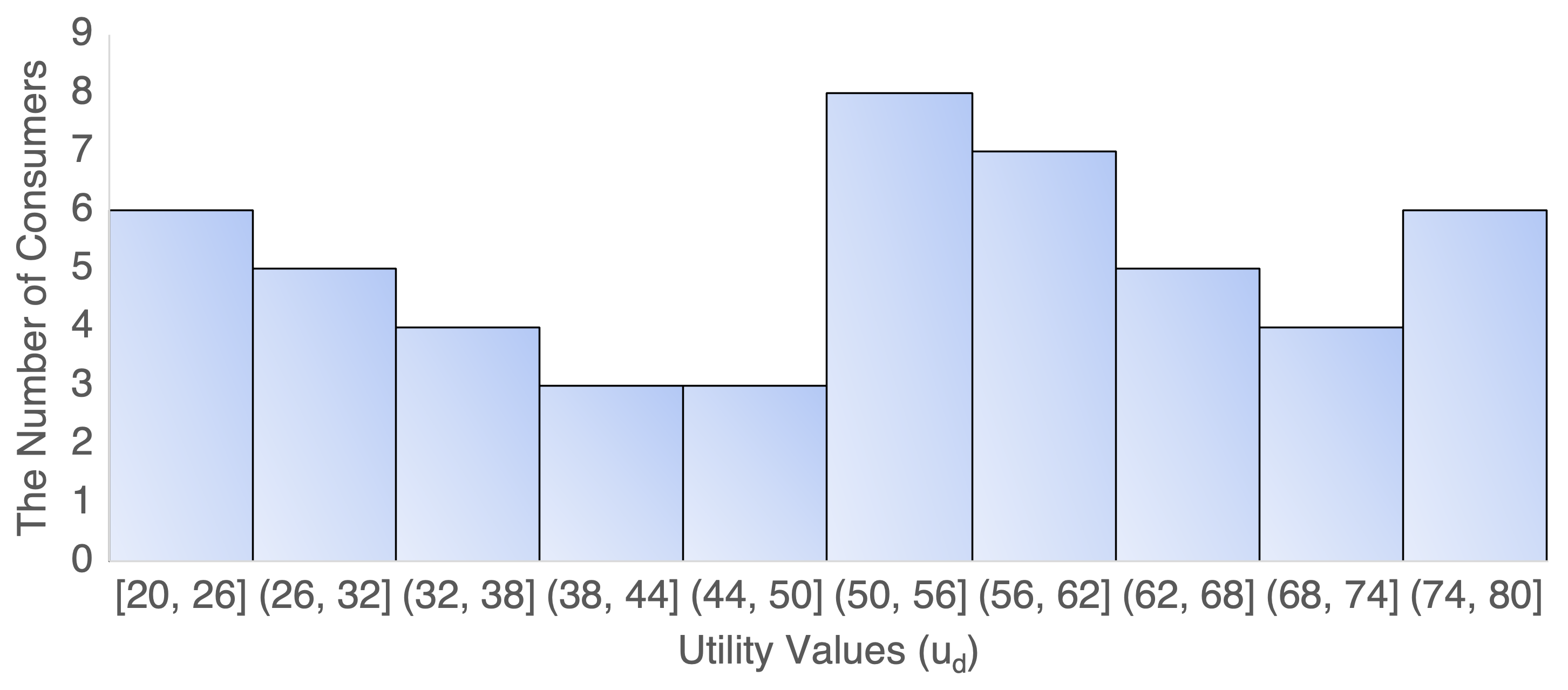}
    \caption{Histrogram showing the utility values $u_d$ of all the consumers in the RTS-GMLC system.}
    \label{figuti}
\end{figure}

\begin{table}[b]
\footnotesize
\centering
\caption{Generator carbon emission intensities.}
\label{tab9}
\begin{tabular}{cccc}
\hline
Type& Carbon Emission Intensity(tons/MWh)\\
\hline
Natural Gas & 0.6042 \\
Oil & 0.7434 \\
Coal & 0.9606\\
Wind, Solar, Hydro &0\\
\hline
\end{tabular}
\end{table}
We use the IEEE RTS-GMLC system \cite{barrows2019ieee} for further numerical analysis. This system has $73$ buses, $158$ generators, and $120$ lines, which are distributed in $3$ regions. For generator fuel types, this system covers natural gas, oil, coal, and renewable sources (solar, wind, and hydro). The emission intensities of different generators are assigned according to the US Department of Energy \cite{emissiondata} and are listed in Table \ref{tab9}. For simplicity, we consider each non-zero load in the system as a consumer, leading to a total of $51$ consumers located at different buses. According to \cite{carbonp}, carbon prices in the world-wide emission trading markets range from $10$ to $100$\$/tons, which can be seen from Fig. \ref{figcp}. Based on these statistics, we assume that consumers will have carbon prices in the range between 
$[10,80]$ in the following experiments. For consumers' demand flexibility, we use the values in the original system as the maximum values and 80\% of these values as the minimum values. The statistics of utility values are shown in Fig. \ref{figuti}.

\subsubsection{Analysis for Different Carbon Costs}
\label{adcb}
We first analyze the impact of different carbon costs on the market clearing outcome, while assuming that all consumers submit non-zero carbon costs to help reduce carbon emissions. We compare cases where the carbon costs of consumers are randomly drawn from a uniform distribution with ranges $[10,20]$, $[10,40]$, $[30,60]$, and $[50,80]$, respectively. We also compare against the carbon-agnostic market clearing models P(2), where all demands are fixed, and P(3), where we have flexible but carbon-agnostic consumers (this formulation is equivalent to our proposed formulation, but with all consumers submitting zero carbon costs). Table \ref{tab11} shows the different components of the objective function: total generation, total carbon emissions, and average carbon emissions. The top rows represent the two carbon-agnostic formulations, and the average carbon cost of the consumer bids increases as we go further down in the rows.
\begin{table*}[!htpb]\footnotesize
\centering
\caption{The impact of different carbon costs on generation dispatch, system emissions, and objective function components.}
\label{tab11}
\begin{tabular}{cccccccc}
\hline
Cases& \begin{tabular}[c]{@{}c@{}}Total Generation\\~[MWh]\end{tabular}  &\begin{tabular}[c]{@{}c@{}}Total Generation \\ Cost [\$]\end{tabular} &\begin{tabular}[c]{@{}c@{}}Total Carbon\\~[tons]\end{tabular}&\begin{tabular}[c]{@{}c@{}}Average Carbon\\~[tons/MWh]\end{tabular}& $u_d^\intercal P_d$ [\$]&$c_{co_2}^\intercal E_d$ [\$]& \begin{tabular}[c]{@{}c@{}}Optimal Objective\\ Values [\$]\end{tabular}\\
\hline
P(\ref{eq123})&8550	&63748&	3001.8&	0.351&457039&	0		&393291\\
P(\ref{eq1234})	&8550	&63748	&3001.8	&0.351&457039&	0		&393291\\
$[10,20]$&8550&64723.4&2866.9	&0.335&457039	&35591.4	&	356724.2\\
$[10,40]$&8550&65468.5&	2804	&0.328&457039&	50088.7	&341481.8\\
$[30,60]$&8263.3	&61675.1&	2594.6&	0.314&448896.9	&99168.8		&288053\\
$[50,80]$&8063.2	&58907	&2473.7	&0.307&439138.2	&143270.5		&236960.7\\
\hline
\end{tabular}
\end{table*}


We first observe that the carbon-agnostic market clearing models P(2) and P(3) lead to similar solutions. This indicates that the utility values $u_d$ of the loads are high enough to cause them to consume at their highest level in the formulation with load flexibility, but no carbon costs. 

Next, we consider the impact of moderate, but non-zero carbon costs in the ranges $[10,20]$ or $[10,40]$ (middle two rows of Table \ref{tab11}). For these carbon costs, we observe a negligible impact on total power generation, indicating that the loads are still consuming the same amount of electricity despite incurring a carbon cost. We observe that the total generation costs increase and the total and average carbon emissions decrease, though not by the same percentages. For the case with carbon costs in the range $[10,40]$, generation cost increases by 2.6\%, while both total and average emissions are decreased by 6.5\%. 
We conclude that when carbon costs are moderate, they can promote the dispatch of cleaner generation sources, while keeping electricity both clean and cheap enough to allow loads to consume at the same level as before. 

Further, we consider the case with higher carbon costs in the ranges $[30,60]$ or $[50,80]$ (lower two rows of Table \ref{tab11}).
As the carbon costs increase to these levels, the total generation decreases. This happens because some consumers are dispatched at a lower consumption than their maximum values, as their carbon costs adjust the utility through the term $u_d^\intercal P_d-c_{co_2}^\intercal E_d$, which becomes too low to support consumption from carbon-intensive or expensive generators. As the total generation decreases, the total generation cost and total carbon emissions decrease even faster. For the highest carbon cost case, the total generation decreased by 5.6\%, while the total generation cost decreased by 7.5\% and emissions by 17.5\%. The average carbon emissions for this case are 12.5\% lower than the carbon-agnostic solutions. 

In summary, we observe that moderate carbon costs can lead to changes in the generation dispatch that increase the dispatch of cleaner generation sources and slightly increase costs, though without impacting total generation or consumption. With higher carbon costs, we observe that carbon emission reductions are achieved both through changes in the generation dispatch and an overall reduction in the total amount of electricity demand. 


\subsubsection{Analysis for Different Proportions of Carbon-Sensitive Loads}
\label{adpc}
We next explore how our results change as the fraction of carbon-sensitive loads, i.e., the fraction of loads that submit non-zero carbon bids, changes. To investigate this, we consider 10 different cases, where the fraction of carbon-sensitive consumers varies from 10\% to 100\% in increments of 10\%. 
Once we have fixed the percentage of carbon-sensitive loads, we randomly generate carbon costs from the higher range of $[30,60]$\$/ton. 
We then define a subset of carbon-sensitive loads to submit their non-zero carbon costs to the market (we assume that all other loads submit zero carbon costs $c_{co_2}=0$). The carbon-sensitive loads are chosen by randomly selecting $x\%$ proportions of total consumers. 
Because of congestion in the system, the locations of carbon-sensitive consumers will impact the results. Therefore, we run our experiment 5 times, each time with a different set of randomly selected carbon-sensitive loads and carbon costs. This allows us to get a sense of how the geographical distributions of the carbon-sensitive loads impact results and makes our analysis less vulnerable to specific choices. 

\begin{figure}
    \centering
    \includegraphics[width = \linewidth]{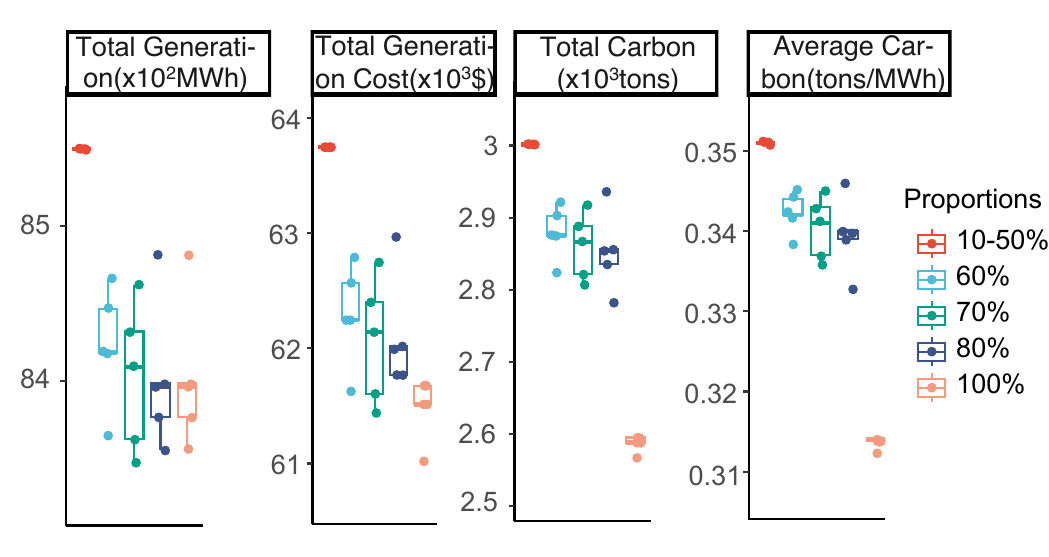}
    \caption{Results for different proportions of carbon-sensitive loads.}
    \label{fig10}
\end{figure}

The results of these experiments are shown in 
Fig. \ref{fig10}, where we see the total generated power (left), total generation cost (middle-left), total carbon emissions (middle-right) and average carbon emissions (right). Each plot shows results for a given proportion of carbon-sensitive loads, with higher proportions further to the right. The dots indicate results from individual simulations, while the boxes indicate average values and the quantiles. 

From Fig. \ref{fig10}, we can see that increasing fractions of carbon-sensitive loads are associated with 
lower total generation, lower total generation cost, lower total carbon emissions, and lower average carbon emissions. Notably, when the proportion of carbon-sensitive loads surpasses 80\%, the influence of load locations on system dynamics becomes less pronounced, 
especially for total and average carbon emission values. Conversely, when the proportion of carbon-sensitive loads is below 50\%, the carbon-sensitive loads have only limited impact on the generation dispatch, as indicated by the cluster of red points in the graph. At intermediate 60\%, 70\% and 80\% proportions of carbon-sensitive loads, the spatial distribution of these loads significantly impacts market clearing outcomes, illustrated by the wider quantiles (indicated by a larger box). 
In conclusion, our findings suggest that our proposed market clearing approach requires a critical mass of carbon-sensitive loads to effectively reduce carbon emissions. Moreover, when the proportion of carbon-sensitive loads falls within the mid-range (60\%-80\% in our study), the spatial distribution of these loads emerges as a crucial factor influencing the performance outcomes. 



\subsubsection{Benchmarking against the Carbon Flow Model} 
\begin{figure}[!htpb]
    \centering
    \includegraphics[width = \linewidth]{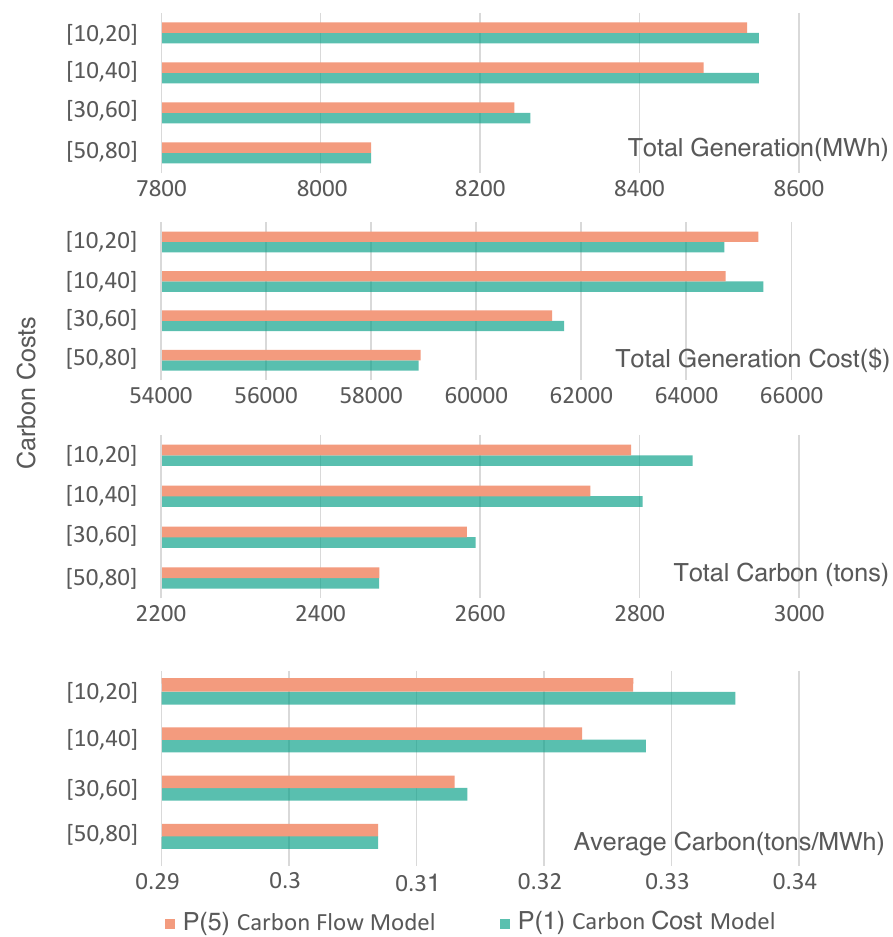}
    \caption{Comparison  between carbon flow model and carbon cost model (different carbon costs).}
    \label{fig11}
\end{figure}

We next seek to compare our results to those achieved if we use the carbon flow method for tracing power from generator to load instead of our proposed optimal allocation approach. 
We first compare results for the case where all loads are carbon sensitive, but we vary carbon costs. To do this, we run the carbon flow model with the same input data as we provided to the proposed model in Section \ref{adcb}. The results for both methods are shown in Fig. \ref{fig11}, with red bars indicating results from the carbon flow model and green bars indicating results from the proposed carbon cost model.

In Fig. \ref{fig11},
we first observe that the carbon flow model leads to consistently lower total generation than the proposed carbon cost model (top subplot). This indicates that a larger number of loads are consuming less due to higher carbon costs in the carbon flow model. The total generation cost (second subplot) is mostly lower as well, which is probably largely due to the reduction in load. However, in the case with the lowest carbon costs $[10,20]$, the proposed model achieves lower total generation costs despite serving a larger amount of load. The total carbon emissions (third subplot) and average carbon emissions (fourth subplot) are consistently lower for the carbon flow model as compared with the proposed model.
Interestingly, the solutions of the carbon flow and the proposed models largely coincide for the highest carbon cost. More analysis and additional experiments would be needed to assess whether this is a coincidence or a trend. 

Overall, we expect that the objective values will be lower with the carbon flow model, since this model is more restrictive in how it allocates generation to loads. The fact that more loads are consuming less energy in the carbon flow model is likely due to the fact that generation from the lowest carbon generators cannot directly be allocated to the loads with the highest carbon cost, e.g., if they are far apart in the network. The proposed model does not face these restrictions and thus is able to serve more load, albeit at the expense of higher generation costs and emissions.





\begin{figure}[!htpb]
\centering
\subfigure[Total generation comparison.]{
\includegraphics[width=\linewidth]{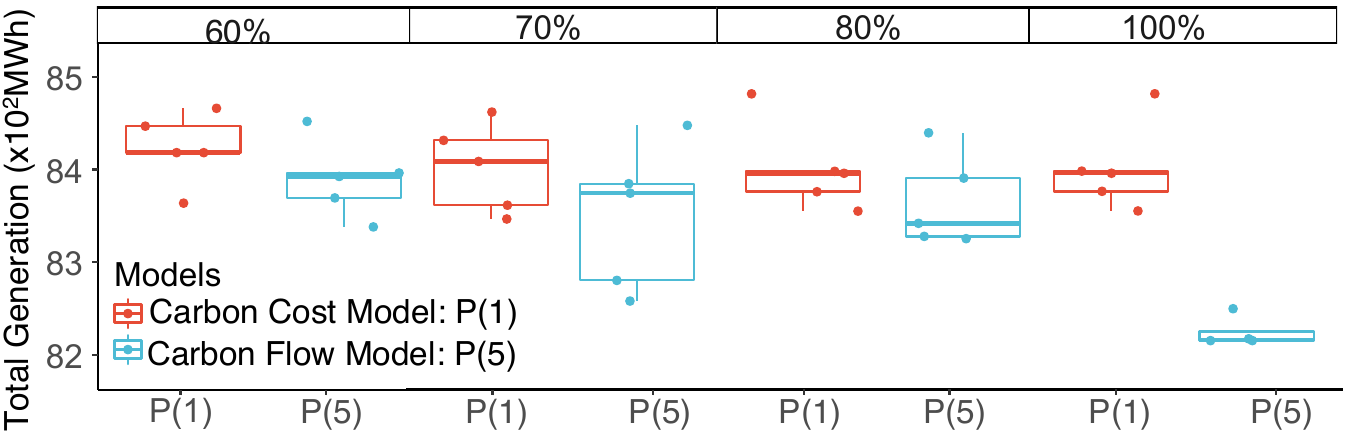}
}

\subfigure[Total generation cost comparison.]{
\includegraphics[width=\linewidth]{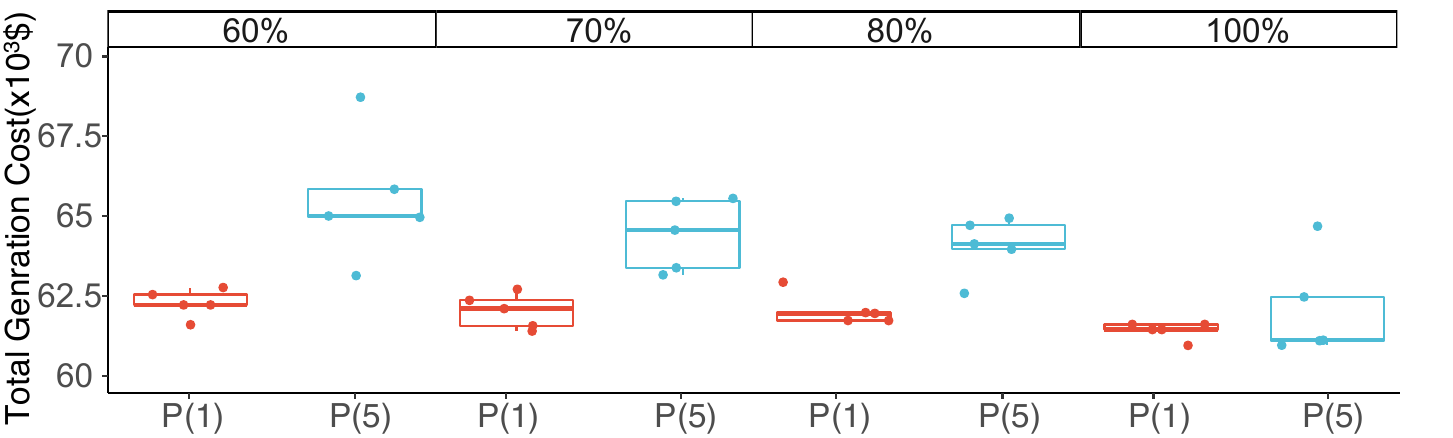}
}

\subfigure[Total carbon emission comparison.]{
\includegraphics[width=\linewidth]{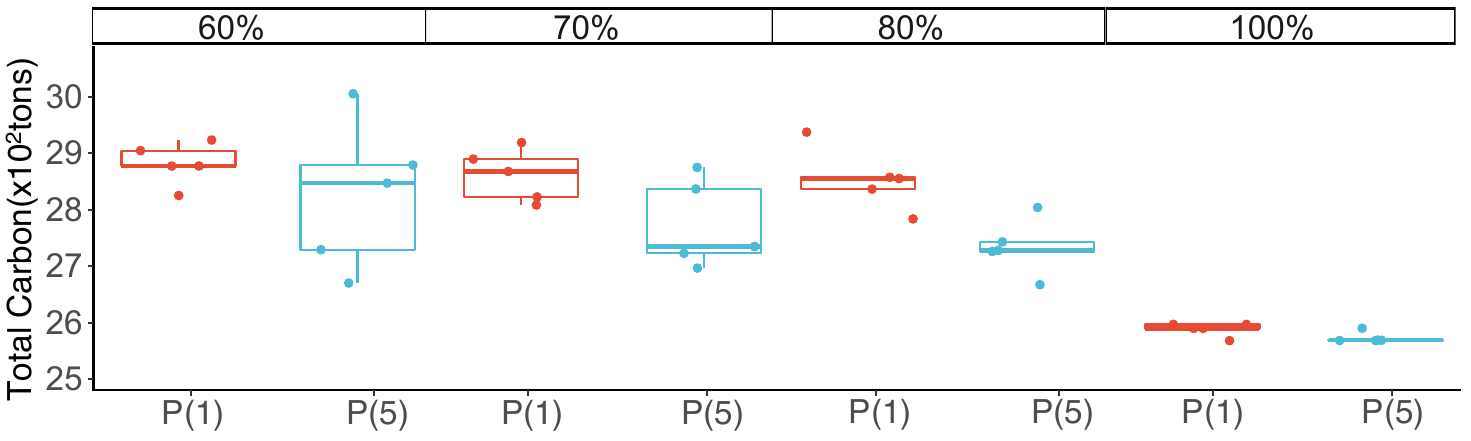}
}

\subfigure[Average carbon emission comparison.]{
\includegraphics[width=\linewidth]{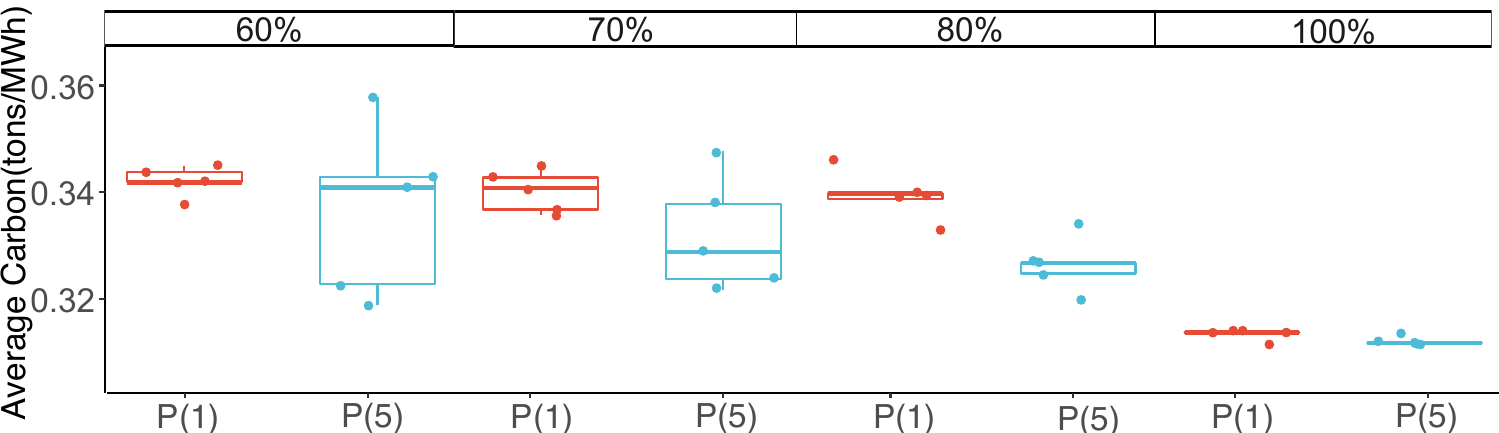}
}
\caption{Comparison  between carbon flow model and carbon cost model (different proportions of carbon-sensitive loads). }
\label{fig12}
\end{figure}

We next compare results from the proposed and carbon flow models as we change the level of carbon-sensitive loads. We do this by repeating the experiments from Section \ref{adpc} with the carbon flow model, using the same input data. The results are shown in Fig. \ref{fig12}, with total generation (top), total generation cost (second), total carbon emissions (third) and average carbon emissions (fourth). We observe that on average, the carbon flow model leads to lower total generation, indicating a more significant reduction in electricity consumption among carbon-sensitive loads. However, for the 60\%, 70\% and 80\% cases, the reduction in load is accompanied by an increase in total generation cost, indicating a significant rise in the average cost of generating power. We further observe that the carbon flow model leads to more significant reductions in carbon emissions.
However, we also observe that there is greater variability in the results of the carbon flow model (indicated by taller boxes), indicating that the location of carbon-sensitive loads matters more in the carbon flow model.



\section{Conclusions}
\label{conclusion}
This paper proposes a novel electricity market clearing model that incorporates the allocation of carbon emissions to consumers and allows consumers to reveal their preferences for avoiding carbon emissions by submitting information about their carbon costs. 
The proposed market clearing formulation provides a new perspective on how to achieve carbon emission reductions. By allocating carbon emission responsibility internally in the market clearing and accounting for consumer carbon costs as part of the price formation,  the optimal generation dispatch is determined based on both generation cost, consumer utility, and consumer carbon costs. 

Our case study demonstrates that the proposed market clearing formulation can achieve carbon emission reductions in two ways. First, the introduction of consumer-based carbon costs can \emph{change the optimal generation dispatch} to promote the use of cleaner, low-carbon generation resources \emph{without reducing consumption} by carbon-sensitive loads. Second, if the carbon costs are high enough relative to the consumer utility and cost of generation, some loads will \emph{reduce their consumption}, leading to further reductions in carbon emissions. 

Further, our analysis indicates that 
the carbon cost model achieves the greatest carbon emission reductions when the proportion of carbon-sensitive loads in a given system exceeds a certain threshold. As the proportion of carbon-sensitive, the carbon reductions achieved by the model are less sensitive to the specific locations of carbon-sensitive loads. 
We also show that the carbon flow model has a more “strict" definition of power flow tracking, and thus less flexibility in the allocation of carbon emissions to loads. As a result, carbon-sensitive loads are typically allocated a larger amount of carbon emissions and are more likely to reduce their consumption.

The conclusions drawn from Section \ref{ns} pertain specifically to the case we study and further work is required to assess the efficacy of the carbon cost model across broader case studies. Nonetheless, 
our results indicate that the carbon bidding model can be a valuable tool for consumers who would like their carbon costs to be accounted for in the electricity market clearing. It is also clear that the proposed market clearing can achieve carbon reductions not only by reducing or shifting demand, but also has a direct impact on the generation dispatch. 


In future work, we plan to expand our analysis to consider additional methods for allocating generation to loads. We also wish to analyze whether our proposed carbon cost strategy aligns individual optimality with social optimality, and perform more thorough analysis for the reasons of carbon emission reduction in different models. We also hope to supplement our analysis with more extensive numerical analysis and more realistic power system test cases.

\bibliographystyle{ieeetr}
\bibliography{ref.bib}


\end{document}